\documentclass{aastex63}
\usepackage{graphicx}
\received{June 26, 2020}
\revised{July 9, 2020}
\accepted{July 10, 2020}
\published{August 13, 2020}
\submitjournal{ApJ}

\begin{document} 
\title{Spatially Resolved Ultraviolet Spectroscopy of the
  Great Dimming of Betelgeuse}
\shorttitle{Betelgeuse Great Dimming}
\keywords{M supergiant stars; stars: Individual: Betelgeuse; stars: chromosphere - stars: mass loss - spectra: ultraviolet}
 
\author[0000-0002-8985-8489]   {Andrea K. Dupree}
\affiliation{Center for Astrophysics $|$ Harvard \& Smithsonian, 60 Garden Street, MS-15, Cambridge, MA 02138, USA}
\email{adupree@cfa.harvard.edu}
\author[0000-0002-6192-6494]{Klaus G. Strassmeier}
\affiliation{Leibniz-Institut f\"ur Astrophysik Potsdam (AIP), Germany}
\author[0000-0002-3728-8082]{Lynn D. Matthews}
\affiliation{Massachusetts Institute of Technology, Haystack Observatory, 99 Millstone Road, Westford, MA 01886 USA}
\author[0000-0002-2554-1351]{Han Uitenbroek}
\affiliation{National Solar Observatory, Boulder, CO 80303 USA}
\author{Thomas Calderwood}
\affiliation{American Association of Variable Star Observers, 49 Bay State Road, Cambridge, MA 02138}
\author {Thomas Granzer}
\affiliation{Leibniz-Institut f\"ur Astrophysik Potsdam (AIP), Germany}
\author [0000-0002-4263-2650]{Edward F. Guinan}
\affiliation{Astrophysics and Planetary Science Department, Villanova University, Villanova, PA 19085, USA}
\author [0000-0002-1640-6772]{Reimar Leike}
\affiliation{Max Planck Institute for Astrophysics, Karl-Schwarzschildstrasse 1, 85748 Garching, Germany, and Ludwig-Maximilians-Universita\`at, Geschwister-Scholl Platz 1,80539 Munich, Germany}
\author[0000-0002-7540-999X]{Miguel Montarg\`es}
\affiliation{Institute of Astronomy, KU Leuven, Celestinenlaan 200D B2401, 3001 Leuven, Belgium}
\author[0000-0002-3880-2450]{Anita M. S. Richards}
\affiliation{Jodrell Bank Centre for Astrophysics, University of Manchester, M13 9PL, Manchester UK}
\author{Richard Wasatonic}
\affiliation{Astrophysics and Planetary Science Department, Villanova University, Villanova, PA 19085 USA}
\author{Michael Weber}
\affiliation{Leibniz-Institut f\"ur Astrophysik Potsdam (AIP), Germany}
\email {others}
\begin{abstract}
The bright supergiant, Betelgeuse (Alpha Orionis, HD 39801) experienced
a visual dimming during 2019 December  and the first quarter of
2020 reaching an historic minimum 2020 February 7$-$13. During 2019 September-November, prior to the
optical dimming event, the  photosphere
was expanding.  At the same time, spatially resolved
ultraviolet spectra using the Hubble Space Telescope/Space Telescope Imaging Spectrograph
revealed a substantial increase in the ultraviolet spectrum and Mg II
line emission from the chromosphere over the southern hemisphere of the star.  Moreover, the temperature and electron
density inferred from the spectrum and C II diagnostics also increased in this hemisphere.  These changes
happened prior to the Great Dimming Event. Variations 
in  the Mg II k-line  profiles suggest material moved  outwards in response to the passage of a
pulse or acoustic shock from  2019 September through 2019 November.
It appears that this extraordinary outflow of material from the star, likely initiated by convective
photospheric elements, was enhanced by the coincidence with the
outward motions in this phase of the $\sim$400 day pulsation cycle. 
These ultraviolet observations appear to provide the
connecting link between the known large convective cells in the photosphere and
the  mass ejection event that cooled to form the dust cloud in the southern hemisphere
imaged in 2019 December,
and led to the exceptional optical dimming of Betelgeuse in 2020 February.
\end{abstract}
\keywords{Stellar chromospheres (230), Stellar atmosphere (1584), Stellar mass loss (1613), M supergiant
stars (988)}

\section{Introduction}
Betelgeuse, (Alpha Orionis, HD 39801), a bright nearby  M2Iab supergiant
star, has long been known to have modulations in both visible light
and radial velocity with a $\sim$2000 day period, coupled with semiregular
fluctuations varying from 300 to 500 days
\citep{Stothers71, Goldberg84, Guinan84, Dupree87, Ridgway13, Chatys19}.  The shorter period of optical 
modulation is believed
to represent radial pulsations in a fundamental or low-overtone
mode, while  the long  period of variation
has been attributed to non-radial gravity modes, binarity, or
magnetic activity \citep{Kiss06, Stothers10, Soszynski14}.

Semiregular photometric variability of Betelgeuse has been
documented for over 180 yr. However,  in 2019 November, the
V magnitude of the star started to fall precipitously (Fig. 1) below its 
expected values \citep{Guinan19}.  During 2020 February  7$-$13,  Betelgeuse 
reached an historic minimum of more than a magnitude dimmer \citep{Guinan20}, than its
typical value of V $\sim$ 0.42 \citep{Johnson66}.
Subsequently, the optical magnitude began to recover. 

This dimming  attracted enormous public attention. Even the casual observer
could see with the naked eye that the star appeared comparable to or fainter than Bellatrix (Gamma Orionis, V=1.64, HD 35468)
markedly changing the appearance of the constellation Orion.  Additionally, Betelgeuse is a candidate to
become a core-collapse Type II supernova \citep{Wheeler17} and suspicions were raised
that its unusual  behavior  could be a harbinger to a massive supernova event.

A visible light image of Betelgeuse, taken with SPHERE using the Very Large Telescope Interferometer (VLTI/SPHERE)
on 2019 December 26 \citep{Montarges20a}, revealed a substantial dimming of the star in the southern
hemisphere causing a change in its apparent shape as compared to a previous image taken
in 2019 January. Observations made on 2020  January 23 \citep{Dharma20} suggest  a decrease in
sub-millimeter flux  when compared to previous epochs that could result from  changes in the photosphere,
either a change in temperature or stellar radius,  the former perhaps resulting from the presence of  cool spots.
However, an optical spectrum obtained on 2020 February 15 suggests a temperature decrease of 100K at most  which
cannot explain the dimming event \citep{Levesque20}.  Possibly the asymmetric appearance resulted from
dust forming in the atmosphere.  Differential speckle polarimetry of Betelgeuse \citep{Safonov20} revealed that
the polarized flux of the circumstellar nebulae around the star remained constant from
2019 October  until 2020 mid-February  when polarization increased.  This strengthens the inferred presence
of a dust cloud.

Here we report spatially resolved ultraviolet (UV) spectroscopy obtained with the Space Telescope Imaging
Spectrograph (STIS)  on the Hubble Space Telescope (HST) spanning the
Great Dimming Event as well as times prior and subsequent to the Event.  We first review the 
radial-velocity measures followed by the flux changes found in the  spatially resolved Mg II emission  and the UV spectrum.
Electron-density diagnostics from C II are evaluated.  Changes in the Mg II k-line profiles are compared to models
of pulsations or acoustic shocks passing through a stellar atmosphere.  And we consider a scenario not associated
with the star to understand the dimming, but resulting from  the presence of interstellar dust clouds. 

%%%FIGURE 1
\begin{figure}[ht!]
\begin{center}
\includegraphics[scale=0.25]{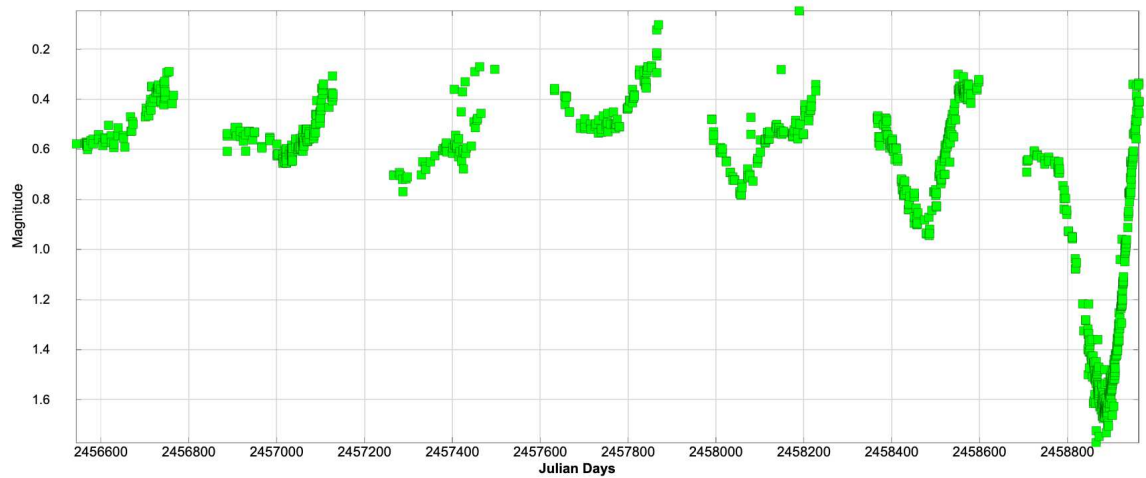}
\caption{V magnitude measures from the Database of the American Association
  of Variable Star Observers (AAVSO, \citealt{Kafka20}). This x-axis begins on 2013 July 26 and
  ends on 2020 May 30  - a span of almost 7 yr.}
\end{center}
\end{figure}
%%%Figure 2
\begin{figure}
\begin{center}
  \includegraphics[angle=90, scale=0.35]{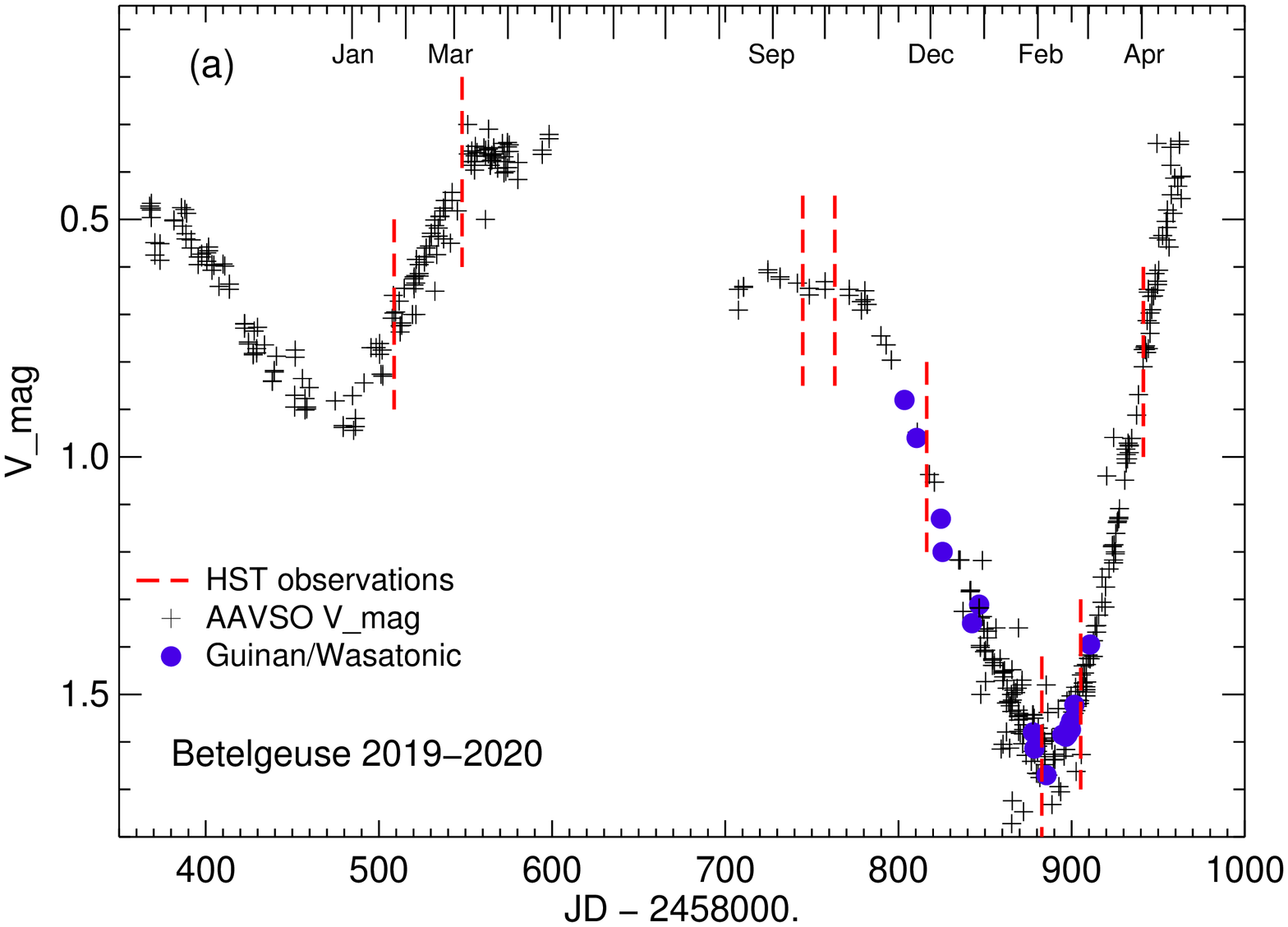}
  \includegraphics[angle=90, scale=0.35]{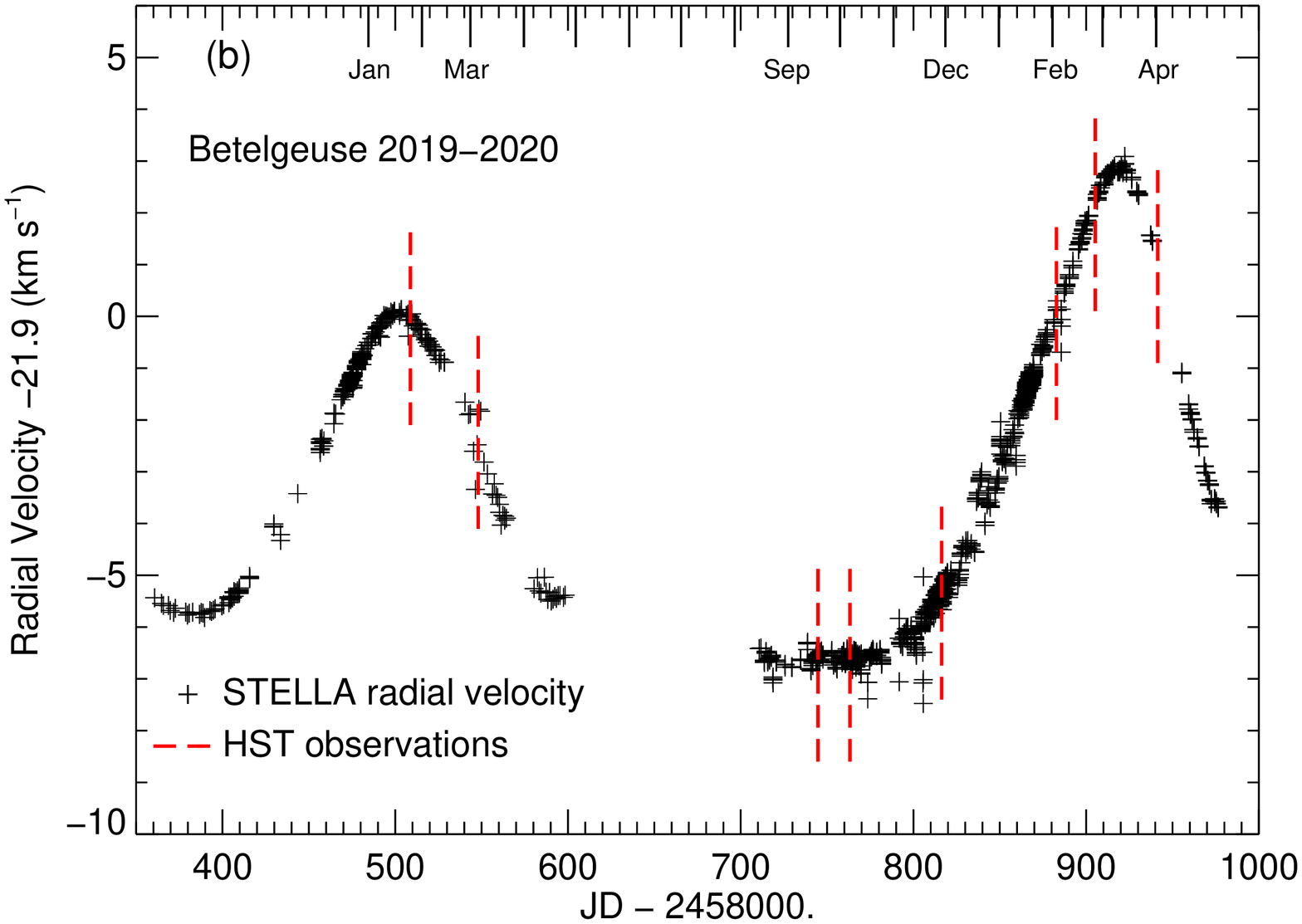}
\end{center}
\caption{{\it Left panel:} V magnitudes from the AAVSO database  \citep{Kafka20} 
and ATel reports \citep{Guinan19, Guinan20}. {\it Right panel:} photospheric radial velocity measures
from STELLA echelle spectra (T. Granzer et al. 2020, in preparation) relative to the average photospheric
velocity of 21.9 km s$^{-1}$ \citep{Famaey05}.
In both figures, the dates of the HST visits are marked
by  broken red lines.}
\end{figure} 

\newpage

\section{Radial Velocity}
Of particular import is the  radial velocity during the time of the Great Dimming (Fig. 2).
Radial-velocity measurements of the photosphere were made
from optical high-resolution spectra obtained with the fiber-fed STELLA Echelle Spectrograph (SES)
mounted on the robotic 1.2m STELLA-II telescope located at Izana Observatory, Tenerife,
Spain \citep{Strassmeier04}.  Beginning, in 2019 January, contemporaneously
with the start of the HST program, the photosphere began accelerating 
outwards from a value near its average velocity,
reaching in 2019 September$-$October, a maximum outflow velocity  of $\sim -$7 km s$^{-1}$ relative to the average.
Outflowing photospheric material was present for almost one year,  from 2019 January
to 2019 November.   At the time of  the HST visit on 2019 November 28,  the photospheric velocity had reversed and inflow 
continued until 2020 mid-March.  By  2020 April,  the photosphere  was moving outward again.

\section{HST Observations}
Time on HST was awarded  for a  program of spatially resolved
UV spectroscopy of Betelgeuse with four visits to the star each year over three Guest Observer
cycles \citep{Dupree18}.   The UV observations are augmented
with multifrequency observations in the radio, infrared, and optical bands to investigate
the process of mass loss from the supergiant originating  in the photosphere and
continuing to the circumstellar environment.  These will be reported elsewhere.
The historic dimming of the
star occurred at precisely an opportune time for the scheduled HST observations during the first two HST cycles.
Figure  2 displays the time of
the HST observations with respect to the optical light curve.  The first visits in 2019 January and  March,
occurred during times of typical photometric behavior and provide  baseline spectra.    Beginning in 2019 September, the HST
visits captured the optical fading with  visits continuing through to 2020 February, followed by the photometric recovery
in late 2020 February, and 2020 April 1 (Table 1).

\floattable
\begin{deluxetable}{lccr}
\tablecaption{STIS Observations of Betelgeuse \label{tab:data}}
\tablecolumns{4}
\tablenum{1}
\tablewidth{0pt}
\tablehead{
\colhead{UT\tablenotemark{a}} &
\colhead{Julian Date\tablenotemark{a}} &
\colhead{Position Angle\tablenotemark{b}} & \colhead{HST Data Set}  \\
\colhead{(YYYY-mm-dd)} & \colhead{(-2,458,000)} &
\colhead{(deg E of N)} 
}
\startdata
2019 Jan 25 & 508.80 & $+$35.1 & ODXG01   \\  
2019 Mar 5 & 548.01 & $+$40.5 & ODXG02  \\ 
2019 Sep 18 & 744.76 & $-$136.5 & ODXG07 \\
2019 Oct 6 & 763.25 & $-$131.5 & ODXG08  \\
2019 Nov 28 & 816.32 & $-$102.4& OE1I01  \\
2020 Feb 3 & 882.82 & $+$35.3 & OE1I52\\
2020 Feb 25 & 905.26 & $+$38.1 & OE1I03\\
2020 Apr 1 & 941.48 & $+$50.6 &OE1I04\\
\enddata
\tablenotetext{a}{At start of first exposure.}
\tablenotetext{b}{Position angle of the long axis of the aperture.}

\end{deluxetable}

%%%%Fig. 3
\begin{figure}
\begin{center}
\includegraphics[scale=0.5]{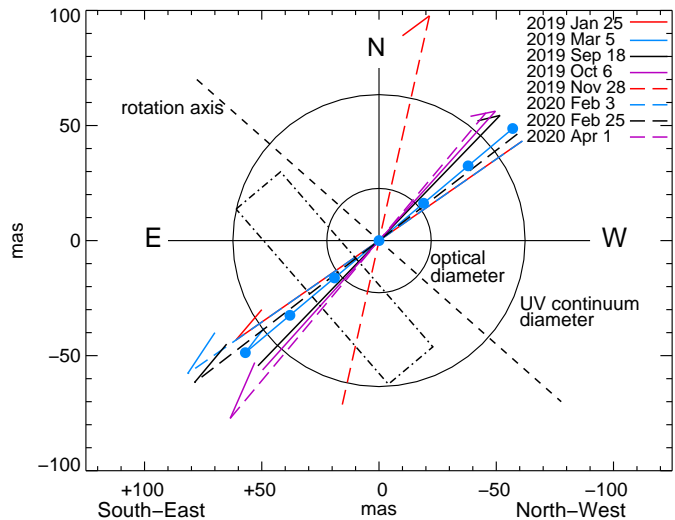}
\caption{Location of STIS pointings for  spectral scans across Betelgeuse. The colored lines mark
  the path of the aperture center; the arrow indicates the direction of 
  motion for each visit. The filled blue circles illustrate the seven pointing
  positions for 2019 March. The others are similar.  The STIS aperture is orthogonal to each line marking the aperture path.
  Spectra were obtained at seven positions ($-$75 mas....$+$75 mas) 2019 January 25-October 6
  and at eight positions during the Cycle 27 visits: 2019 November 28-2020 April 1 ($-$75 mas ... $+$100 mas).
  The optical/near IR  diameter ($\sim$44 mas,  
  \citealt{Haubois09, Montarges16}) and the UV-continuum diameter ($\sim$125 mas, \citealt{Gilliland96})
  are marked. The rotation axis of
  48$^\circ$ (E of N) is denoted by a broken line \citep{Kervella18}. UV emission in the
  Mg II lines extends  to at least $\pm$135 mas \citep{Uitenbroek98}. The size of the  STIS aperture (25 $\times$ 100 mas)
  is marked by the rectangle ({\it dotted-dashed line}).}
\end{center}
\end{figure}

Each visit of HST  to Betelgeuse began with a target acquisition, followed
by two peakup maneuvers with the small slit (25 $\times$ 100 mas) using the dispersed
UV spectrum. 
During the spectral observations, the 25$\times$100 mas aperture of 
STIS was placed across the UV disk of Betelgeuse at seven or eight positions (Figure 3).
An offset pattern of $-$75 mas, $-$50 mas, $-$25 mas, centered,
$+$25, $+$50, and  $+$75 mas was executed for subsequent exposures.  
Exposure times for the spectra ranged from  372s at the center position to 1337s or
2813s at the outer positions.
A near-UV  spectrum was obtained at each offset position. The  E230M grating  yields a
resolution of 30,000 and spans 2275$-$3070\AA\ with the 2707\AA\ setting.   
This spectral  region contains the strong
\ion{Mg}{2} chromospheric emission lines and many other weaker
transitions that can trace the  dynamics as well as the
atmospheric  structure.  

During 2019, HST had several problems with acquisition or reacquisition of
guide stars, and the position of the aperture was not known. We do not use those spectra. New 
visits were successful, and are only used here.  Table 1 contains a summary
of the visits and data sets.  In the second HST cycle,
we were also able  to acquire an additional spectrum at an offset position
of $+$100 mas during each visit.  Because the orientation of the spacecraft changes (see 
Table 1), five scans proceed from the northwest to the southeast across the star,
and three others move in approximately the reverse direction.  Figure 3
shows the positions of the aperture center and direction of motion  for each of the eight visits.

\newpage

\section{Chromospheric Mg II Emission}

The Mg II emission lines, 2795.528\AA\ ({\it k-line}) and 2802.705\AA\ ({\it h-line}) provide evidence of 
plasma at chromospheric temperatures (4000K$-$8000K).  
The shapes of the Mg II h- and k-lines in Betelgeuse (Fig. 4) differ from one another, which is unusual among luminous
cool stars \citep{Robinson95}. Modeling of the Mg II profiles in a semiempirical  spherical  atmosphere,
with a  multilevel atom, and non-LTE
formulation \citep{Lobel00} suggests that the intrinsic shape of the k-line should mimic that of the
h-line. Circumstellar absorption from Mn~I (2794.817\AA) and Fe~I (2795.006\AA) can reduce the flux
in the blue component of the k-line,
but it is not at all clear that these lines can reduce the flux sufficiently to match the observed
profiles.  It may be that wind absorption also plays a role, because the k-line would be affected
more than the h-line \citep{Uitenbroek98} due to its larger opacity.

%%%%FIGURE 4
\begin{figure}
  \begin{center}
  \includegraphics[angle=90,scale=0.4]{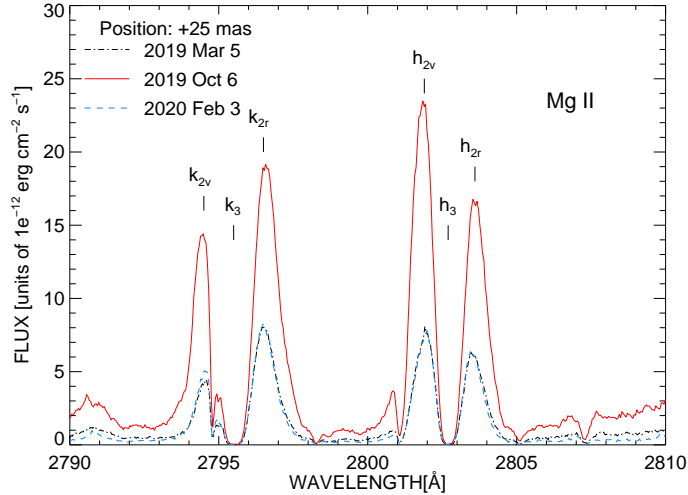}
  \caption{Mg II profiles at the pointing position offset from the center by +25 mas to the southeast.
    Three visits are shown: during the flux maximum on 2019 October 6 as compared
    to the same position on the disk on 2019 March 5 and 2020 February 3. The various components of the 
    emission profile are marked: the central minimum (h$_3$ and k$_3$) and  the emission peaks (h$_2$ and k$_2$).  Here 
    {\it v} denotes the short-wavelength emission and {\it r} refers to the long-wavelength emission of each line.  The
    flux in February has clearly
  returned to the preoutburst condition. Wavelengths are shifted to laboratory values. Narrow absorption
  features due to Mn I (2794.817, 2801.081\AA) and Fe I (2795.066\AA) are present.}
 \end{center}
\end{figure}

The aperture-integrated intensity of each Mg II line as a function of position across
the UV disk is shown in Figure 5a for the eight visits spanning 2019 January 25 through 2020 April 1.
A strong enhancement in the chromospheric emission, by factors of two to four appeared in the southeast quadrant
of the star during  the three visits  spanning 2019 September 18 through 2019 November 28. The flattened
image of the UV spectrum on the MAMA detector (the files with extension {\it flt}) at the brightest
position ($+$25  mas) during
  the outburst (September to November) appears
  symmetric in the cross-dispersion direction\footnote{The cross-dispersion direction is parallel to the long
    axis of the aperture.}, mimicking the known response of the point-spread function.
  This  suggests that the source of the excess emission was not positioned
  at one or the other end of the aperture, but reasonably centered in the southeast.  Following the enhancement 
event, the Mg II flux observed on 2020  February 3 had returned to a level consistent with that
in 2019 January  and 2019 March.  During the `normal' times for the ultraviolet (2019 January$-$2019 March  and
2020 February$-$2020 April), the spatial
distribution of the flux appears reversed, becoming
stronger towards the northwest than towards the southeast.  But note that the optical
V magnitude was historically faint during early 2020 February.

%%%FIGURE 5
\begin{figure}
 
    \includegraphics[angle=90, scale=0.4]{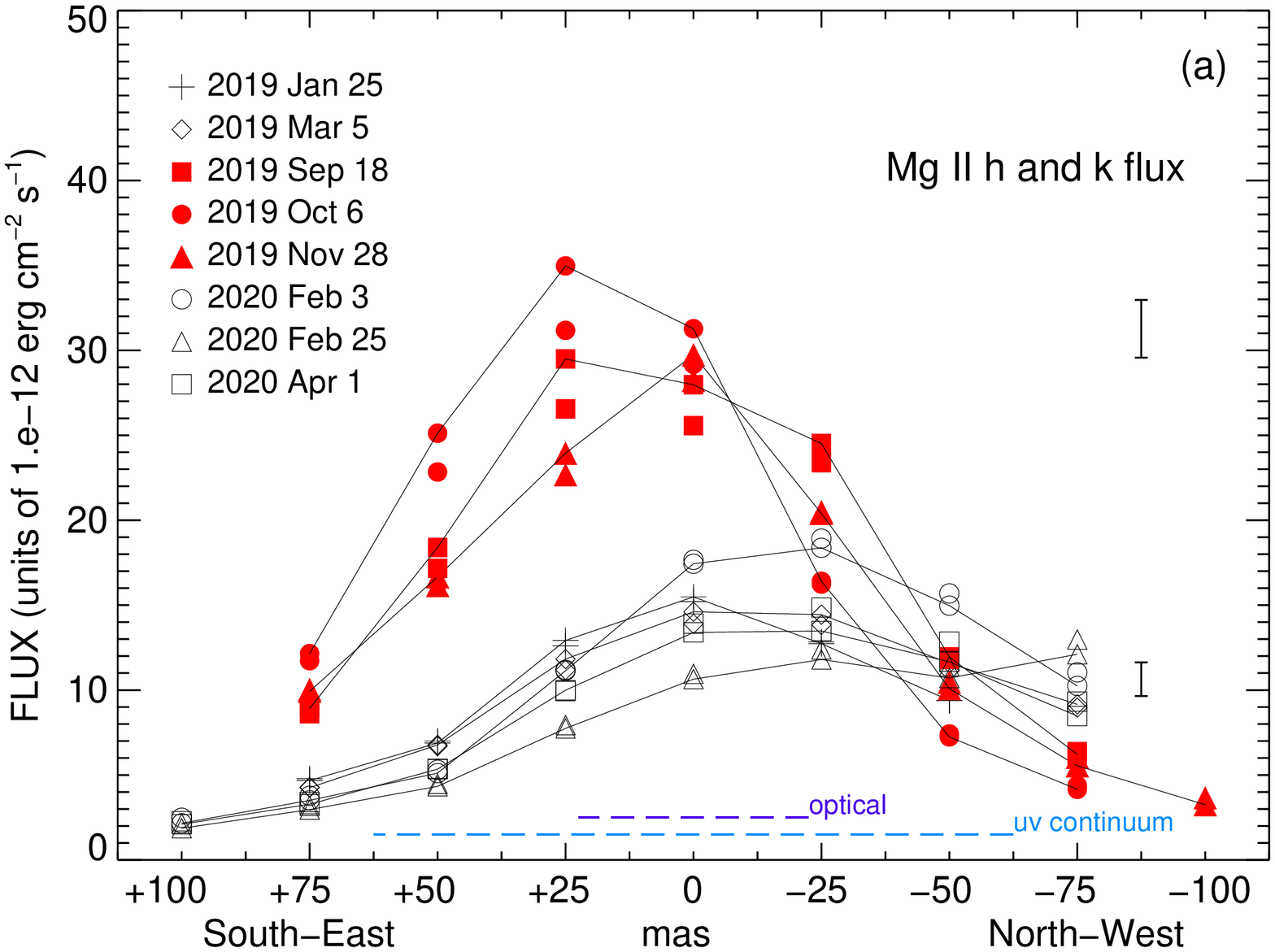}
    \includegraphics[angle=90,scale=0.4]{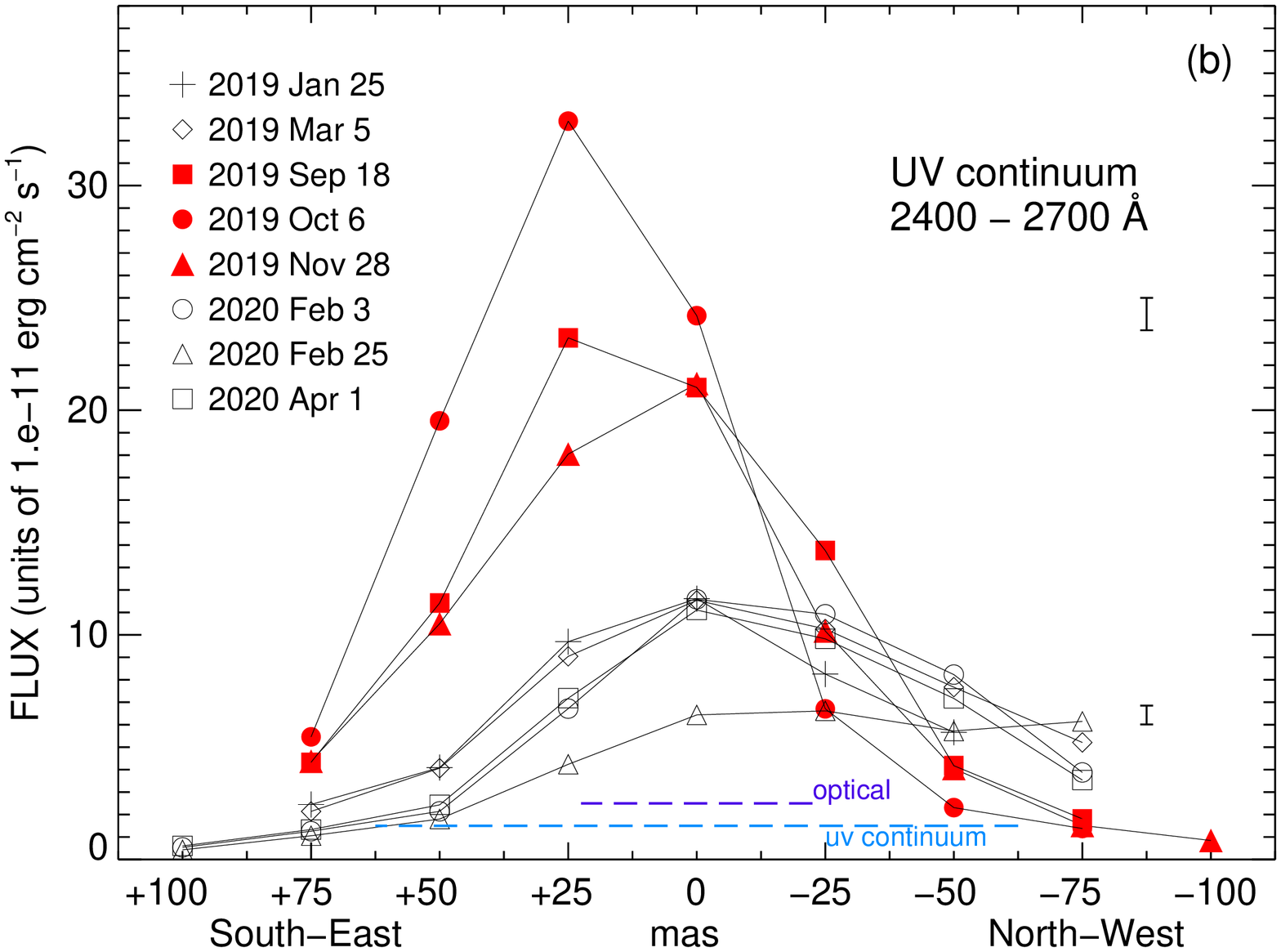}
    \caption{{\it Left panel:} The spatially resolved total flux in Mg II h and k emission  as a function of 
      offset position.  Fluxes for
the h-lines are connected by a solid line.  The approximate extent of
the optical- and UV-continuum diameters with respect to the STIS pointings is marked as
shown  in Fig. 3. Measurements
during 2019 September$-$November, exhibiting substantially increased emission, are marked by red filled symbols.
    {\it Right panel:} total flux in the near-UV spectrum between 2400-2700\AA.}
\end{figure}

The line fluxes differ from those recorded by \citet{Dupree87} using IUE.
Particularly intriguing is the ratio of the total line fluxes:  k-line to h-line.  
During 1984-1986, the disk-integrated  k-line flux was consistently equal to or
stronger than the h-line flux by
factors ranging from 1.0$-$1.4. However the spatially resolved STIS spectra,   
beginning in 2019 March  and extending through 2019 October  exhibit a k-line that has a lower flux than the
h-line. The largest decrease in the spatially resolved Mg II k flux,
namely 11\% (leading to  a  k/h ratio of 0.89) occurs in the southern
hemisphere (position +25mas)  in 2019 October. By 2020 February 3,  the lines revert back to their `normal' ratio:
namely the flux in the k-line becomes comparable (to within 3\%) to  that in the
h-line.\footnote{Creation of a disk-integrated flux in the h and k-lines,
  by summation over all the STIS pointings at each visit produces a ratio of the k-line to h-line flux that
  is consistent with 1.0 to within $\pm$6\%. The disk-integrated ratio from the  2019 October
  pointing is the lowest at 0.94.}
We speculate that increased absorption, either from the circumstellar Mn I and Fe I
features or from outflowing material or both, could cause this  weakening of the k-line flux. Alternatively, the
structure of the  chromosphere  changed substantially from earlier times to produce the different flux ratio.

\section{UV Broad-Band Emission}
The near-UV  flux from 2400-2700\AA\ is shown in Figure 5b as a function of position for the eight visits.
An enhancement in the southern hemisphere, similar to that found in the Mg II emission, also occurs in these
broad band observations in 2019 September$-$November.
The flux at the $+$25 mas position, at maximum in 2019 October, is a factor of 3.5 larger than in the
2019 January$-$March  observations.  Contrast this with the enhancement of the h-line of 2.9 and the k-line of a factor of 2.4.
Because the broadband UV is formed closest to the stellar photosphere,  the  Mg II h-line above that,
and the Mg II k-line highest of all in the atmosphere, such a systematic change suggests the presence of a substantial
photospheric phenomenon.  A similar behavior was noted \citep{Uitenbroek98}  in the near-UV spectra
obtained with the larger aperture of the GHRS, which was  accompanied
by a contemporaneous HST Faint Object Camera (FOC) ultraviolet image.  The UV image revealed a single
large unresolved bright area in the southwest quadrant of the disk \citep{Gilliland96}, and
a systematic enhancement in Mg II was present.
Thus we conclude that during the months of 2019 September through November, a similar
bright area was present in the southeast quadrant.  We identify the position of maximum flux in the southeast
because in the majority of offset pointings that position marks the aperture center. Additionally, as noted earlier, the spectrum in the direction of the long axis of the aperture 
appears symmetric across the 4 pixel aperture.

Under typical conditions, the spectrum near 2500\AA\ approximates a blackbody with T$\sim$ 5000K
\citep{Gilliland96}. An enhancement of a factor of 3.5 at maximum, such as occurred in the
south in 2019 October, interpreted as a temperature change above the photosphere, suggests an increase in temperature of
about 600K, to 5600K. This exceeds the enhancement found in 1995 in which the bright area
on the UV disk image was a factor 1.3$-$1.8 brighter than the surrounding disk, and suggested
a temperature differential of $\sim$200K. The event in the fall of 2019 appears to be more energetic.

\newpage

\section{Electron Densities}
Near-UV transitions of C II offer a direct measure of the electron density \citep{Judge98, Harper06}. The ratio of
two transitions, 2325.398 and 2328.122\AA\ between the
configurations $2s^2 2p\ ^2P^0$ and $2s 2p^{2\ 4}P$ is sensitive to electron density.  These lines are not
significantly blended and are easily deconvolved.  The flux ratios 
from the spectra at the center and at  position +25 mas ({\it the pointing with maximum flux in the southeast})
are shown in Figure 6.   While the central position appears consistent with a constant ratio,
the ratio at the +25 mas position is lower in 2019 September$-$November  than prior or subsequent to this time.
This lower value signals   an increase in the electron density.  Atomic calculations  from \citet{Harper06}, assuming a
temperature of 6300K, suggest  
a density increase  by  $\sim$0.5 dex or more during the chromospheric outburst in the fall of 2019 depending
on the optical depth of the lines.  The density
returned to preoutburst value by 2020 February.  Thus the southern
hemisphere displayed a warmer denser plasma during the 2019 September$-$November outburst as compared with  prior and
subsequent times.  A detailed examination of other members of the multiplet and different offset positions is
underway.

\begin{figure}
  \begin{center}
  \includegraphics[scale=0.55]{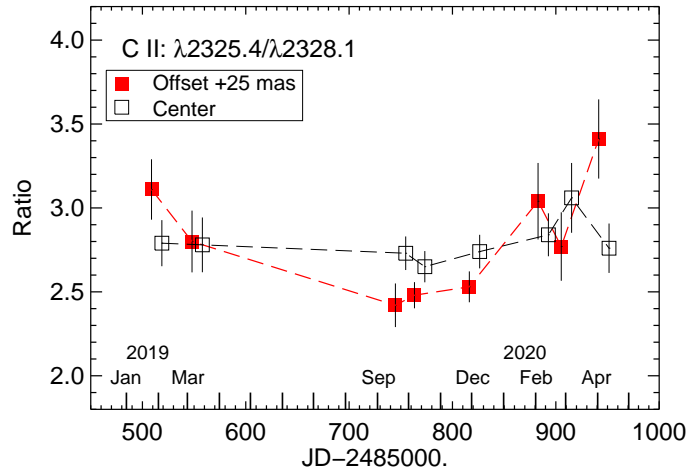}
  \caption {Ratio of C II emission lines: 2325.4\AA/2328.1\AA\ at the offset position of +25 mas and disk
    center. This   +25 mas 
    position on the southeast quadrant of the disk corresponds to the maximum in the Mg II flux and continuum
    as shown in Fig. 5.  Error bars on the ratio derive from the errors in the individual line flux measurement as provided
    by the CALSTIS pipeline reduction.  The central position appears consistent with a constant electron density.  The
  +25 mas offset position suggests an increase in the electron density 2019 September$-$November.}
  \end{center}
\end{figure}

\section{Temporal Changes in Profiles}

The near-UV spectrum contains  transitions that can indicate atmospheric dynamics by the shape and changing
appearance  of the line profiles.  Chromospheric Ca II and Mg II lines in cool stars have been scrutinized 
and modeled intensely, particularly with respect to high-resolution
solar observations \citep{Leenaarts13}.
Diagnostics based on line strengths, asymmetries, or Doppler shifts in the Sun cannot be easily transferred
to the spectrum of a supergiant, partly because of narrow absorption features in the Betelgeuse Mg II 
spectrum (Fig. 4)  that cross the
line profile and the broad deep  central reversal (h$_3$ and k$_3$).  And of course,  atmospheric
conditions and radiative transfer effects in a luminous supergiant star differ substantially from those of a dwarf star.
However, several  models of the Sun consider the simple response of the Mg II k-line profile to the  passage of a pulse or
an acoustic shock through the atmosphere  \citep{Gouttebroze80, Carlsson92}.  These models
demonstrate  that the pulse produces enhanced
emission on the short-wavelength side of the line (k$_{2v}$) due to the increased temperature
and density of the material moving outward. And in fact, the observed  ratio, k$_{2v}$/k$_{2r}$ in Betelgeuse
spectra (Fig. 7)  does show such
an increase during 2019 September$-$November, suggesting the presence of an outwardly moving disturbance in the
southeast quadrant of the star. 

%%%%Figure 7
\begin{figure}
  \begin{center}
  \includegraphics[scale=0.55]{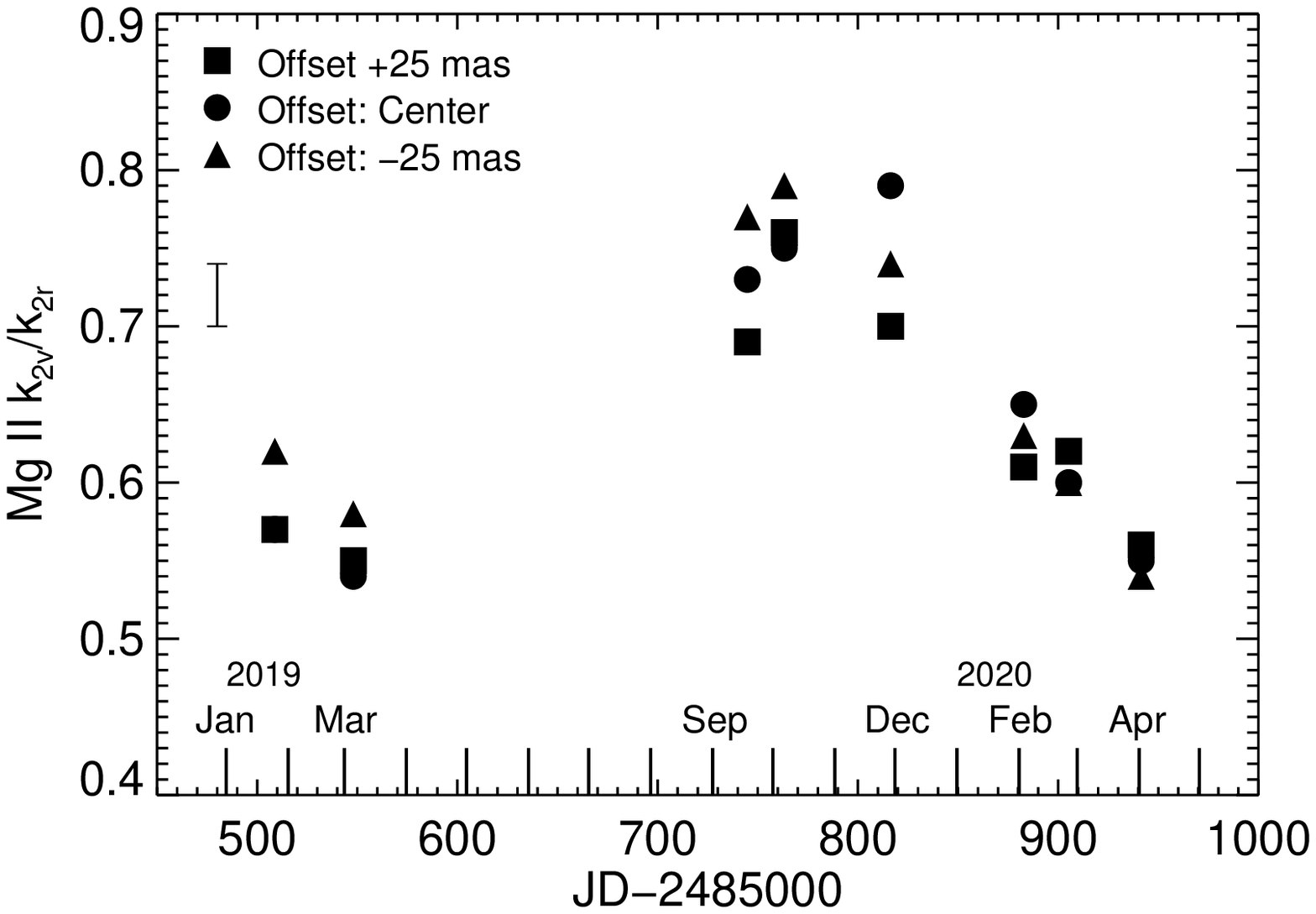}
  \caption{The ratio of the peak flux of Mg II k$_{2v}$/k$_{2r}$ from the STIS pointings on the center
    of the optical disk and a $\pm$25mas offset for the eight visits to Betelgeuse, suggesting the outward motion of
    a pulse of chromospheric plasma  through the atmosphere.  A typical error bar is shown as derived from the HST/CALSTIS
  pipeline.}
  \end{center}
  \end{figure}

The disk-integrated Ca II K-line generally shows a similar profile and  behavior during 2019 and 2020.
Echelle spectra obtained at the Tillinghast Reflector Echelle Spectrograph (TRES) at the Fred Lawrence Whipple Observatory (FLWO),
reveal a similar change in Ca II K-line as the Mg II k-line.  
The short-wavelength emission peak (K$_{2v}$) generally becomes stronger then the long-wavelength emission peak (K$_{2r}$)
beginning in 2019 April 18 and continuing though 2019 October. However, both prior (2019  March 5)  and subsequent to the
chromospheric outburst (2020 January 17), the strength of the peaks is reversed$-$that is, K$_{2v}$ $\leq$ K$_{2r}$. 
This asymmetry suggests that a stable  outflow has returned to the chromosphere.

%%%%FIGURE 8
\begin{figure}[ht!]
\begin{center}
\includegraphics[scale=0.6]{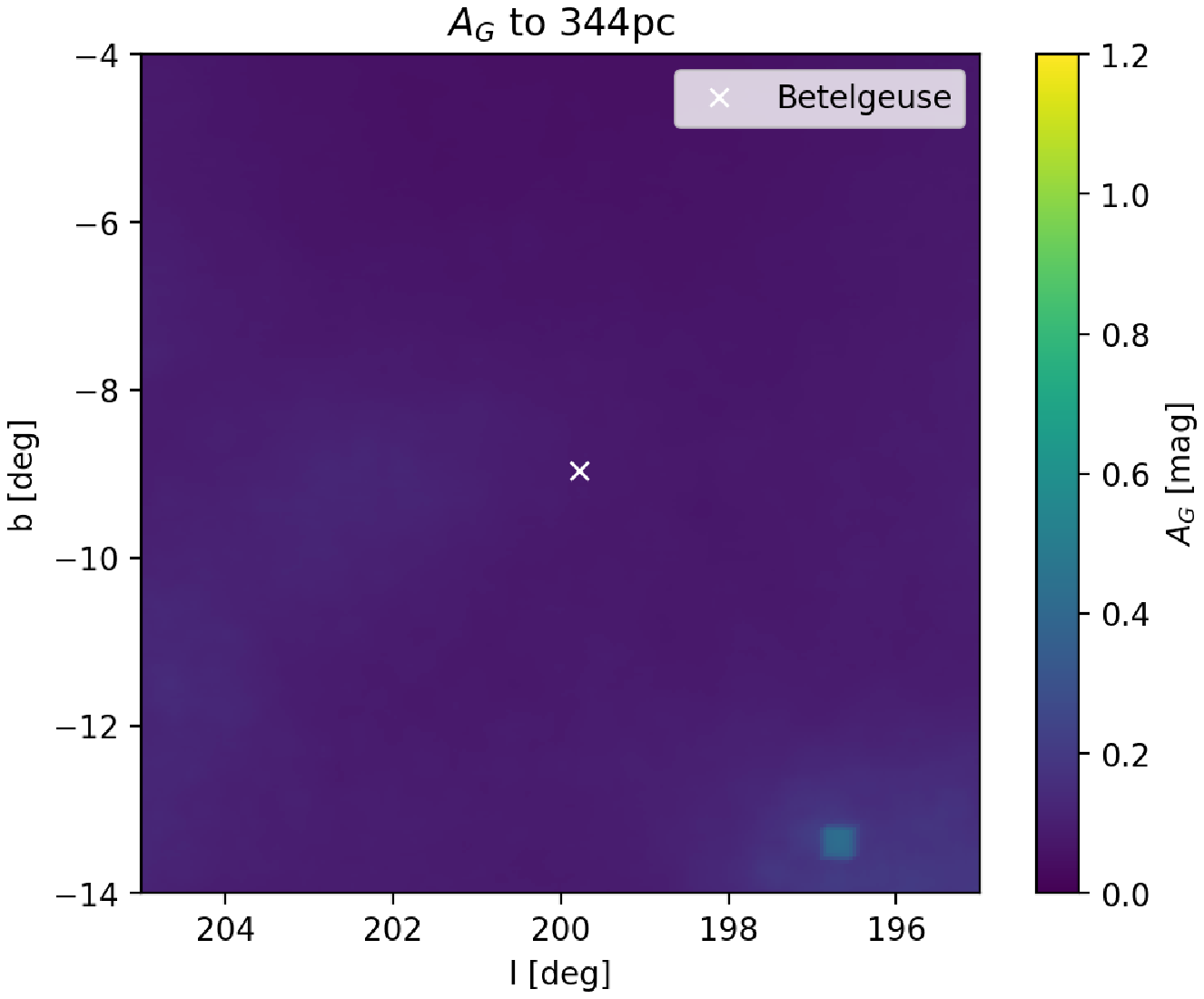}
\caption{Map of the integrated G-band extinction (in magnitudes) in sky projection centered on the
  position of Betelgeuse.  This is calculated at the furthest distance of Betelgeuse, 344 pc, allowed by
  a 2$\sigma$ error on the parallax  \citep{Harper17}.
      No sign of nearby absorbing structures is evident.  Inspection of the standard deviation of the
    extinction models shows it is uniformly less than 0.062 in this region.  The `bright' point
in the lower right corner does not correspond to any stellar source and is believed to be noise.}
\end{center}
\end{figure}

\section{A Transiting Cloud?}
The VLT/SPHERE optical image of Betelgeuse \citep{Montarges20a, Montarges20b} made in 2019 December 
revealed an oddly shaped stellar image displaying a faint southern hemisphere. That image was reminiscent of an earlier image
of the Epsilon  Aur system \citep{Kloppenborg10}. Epsilon Aurigae is an eclipsing binary star system, in which
the primary is an F-type star, and the secondary is a single B5V-type star surrounded by an opaque dust disk.
The dust disk  transited the primary star, causing a large dark region to move across the southeast to
southwest quadrant of the F star over 1 month. Currently Betelgeuse is not believed to  have
a companion.\footnote{\citet{Karovska86}
  reported that Betelgeuse has 2 companions using a new photon-counting camera and speckle
   imaging reconstruction.  However subsequent high resolution imaging has failed to confirm the presence of
   companions \citep{Hebden86, Christou88, Buscher90, Kervella09}.}  The region
surrounding Betelgeuse contains complex bow shocks as revealed by the Herschel satellite \citep{Decin12}.

An alternate scenario  to explain the anomalous image is that Betelgeuse moved behind an interstellar dust cloud
not produced by the star,  that obscured the southern
hemisphere.  Recent three-dimensional mapping of nearby interstellar dust clouds  was constructed by  \citet{Leike20}
by using combined data from Gaia, the Two Micron All Sky Survey, PANSTARRS, and ALLWISE.
They modeled dust densities in clouds up to a distance
of 400 pc with a resolution of 1 pc  amounting to 15 arcminutes  at 222 pc, the distance of Betelgeuse \citep{Harper17}.
Figure 8 shows the expected integrated G-band extinction for a 10 degree field on the sky centered on Betelgeuse.
The  figure displays the extinction at the maximum distance of Betelgeuse indicated by the lower error bound
on the parallax, corresponding to 344 pc. No sign of dust is apparent. Calculation of
the extinction at the distance of the star (222 pc) and the higher error bound on the parallax (164 pc)  produces images that
replicate the image at 344  pc with no indication of any dust features.  Extinction in other directions as displayed in
\citet{Leike20} clearly demonstrates the generally unobscured line of sight to Betelgeuse.  While these observations
cannot exclude the presence of isolated dust features at the scale of 1 pc or below, they do suggest that
Betelgeuse is located in a generally dust-free region.  It does not appear likely that the star moved behind a 
dust cloud.  This confirms the results from \citet{Levesque20}, who show that the 
optical spectrum of Betelgeuse is attenuated more uniformly than would 
be expected from typical extinction by interstellar dust.

\section{Discussion}

These  results   reveal the connection between the photosphere, the chromosphere, and
the subsequent dust production that appeared to obscure Betelgeuse in 2019 December  leading to the 
deep optical dimming in 2020 February.   The UV observations clearly indicate a bright, hot, dense
structure appeared in the upper photosphere and low chromosphere in the southern hemisphere of the star during 2019 September
through November.  This bright structure was present before  the onset of the optical dimming event.
During this time the photosphere was expanding (Fig. 2) following its $\sim$ 420 day pulsation period.
The UV line profiles signaled  the passage of a pressure wave outwards through the atmosphere.
The photosphere and chromosphere of Betelgeuse are well-documented
to exhibit large variable  bright areas most likely arising from large convective cells in the
photosphere \citep{Haubois09, Kervella09, Montarges16}  creating hot regions
 in the chromosphere \citep{Gilliland96, Dupree13}.

We speculate that a convective upflow in the photosphere initiated a major outflow event,  and its effect was
enhanced by the coincidence with outward motion present at this phase of the pulsation cycle, combining
the upward motion of these two distinct physical mechanisms into an extraordinary strong upflow. This
plasma  produced the substantial increase
in UV line and continuum emission 2019 September$-$November.  Eventually the material cooled,
forming the dust that obscured the southern hemisphere
of the star revealed in the VLTI/SPHERE optical images from 2019 December 26 \citep{Montarges20a, Montarges20b}.  The dust opacity
increased rapidly over the next $\sim$ 6 weeks, causing the
exceptional  deep optical dimming in 2020 February by which time the stellar atmosphere had returned to its previous state.

\citet{Kervella18} suggested that 'rogue' convective cells in the photosphere observed in the submillimeter range 
emitted a focused molecular plume that could condense into dust.    
A molecular plume associated with this event should be sought - and would provide added confirmation of
this scenario.

The spatial location of the 2019 chromospheric event reported here is of particular interest.  Previous spatially
resolved spectroscopy  in both the UV \citep{Uitenbroek98} and the submillimeter ranges  \citep{Kervella18} have
given indications of the rotational axis of Betelgeuse.  Based on velocity shifts of photospheric lines
and the presence of a bright spot in the south, \citet{Uitenbroek98} identified the spot with the pole of
the rotation axis, placing the rotation axis at 55$^\circ$ E of N. This is in harmony with polarization measures
and early theoretical models \citep{Asida95}  of the emergence of convective elements in the photosphere.
Moreover, extended 'plumes' have been observed in the southwest from the star \citep{Kervella09}.
ALMA measurements of SiO emission in the submillimeter
region \citep{Kervella18} suggest that the  north pole of the star is directed
toward 48$^\circ\pm$3.5 (E of N).  Again, the presence
of a hot spot, a molecular plume and a pre-existing dust shell toward the north indicated that the north pole was a
source of a mass outflow event. Association with the rotation pole may indicate the presence and effect
of a dipole magnetic field.  However, the source of this 2019 chromospheric event does not appear
aligned with either rotational pole of Betelgeuse. This brightening  occurred in the southeast quadrant,
and at a location different from the position of either stellar pole.  Other bright spots in
the photosphere \citep{Haubois09},
chromosphere \citep{Dupree13}, and extended atmosphere \citep{OGorman17} have occurred on the disk and 
away from the polar regions. Previous UV spatially resolved spectra \citep{Lobel02} revealed the presence
of large scale global nonradial oscillations in the chromosphere of Betelgeuse. It seems likely that
large photospheric convective cells, chromospheric events,  and
subsequent mass outflow exhibit no preference  to originate at the rotation poles.

The Great Dimming of Betelgeuse may be similar  
to the R Cor Borealis phenomenon.  R Cor Bor objects are post-AGB supergiants that undergo substantial dust-formation
episodes and rapid fading of optical light.  The fading in the optical can occur over
several days or weeks \citep{Clayton13}.   Observations suggest the dust forms in the atmosphere and then is
radiatively accelerated to high velocities (220 km s$^{-1}$).   The chromospheric outflow as measured in the
He~I 10830\AA\ transition appears during the decline when radiation pressure accelerates the dust and gas.  Unfortunately,
  Betelgeuse does not exhibit this He I transition, and its radiation field is weaker than the  warmer R Cor Bor objects.
  We have not yet measured  any such rapid acceleration in the atmosphere of Betelgeuse.  However, the time scale
  for the inferred dust formation in Betelgeuse, namely 1-2 months,  is comparable to the R Cor Bor stars.

We are fortunate that  Betelgeuse is relatively close by so that features on the star and in its extended atmosphere can
be resolved to  probe the mass-loss process and the circumstellar environment of a supergiant. We anticipate
the next minimum in $\sim$420 days from 2020 February  which will happen in 2021 April, unfortunately as Betelgeuse moves
into the daylight hours for observations from Earth. However, the presence of spacecraft, such as STEREO,
at different positions in the Earth's orbit, and landers on Mars, may fortunately  allow us to follow the next minimum.

We acknowledge with thanks the variable star observations from the {\it AAVSO International Database} contributed by observers
worldwide and used in this research. Results are based partly on data obtained with
the STELLA robotic telescope in Tenerife, an AIP facility jointly operated by the AIP and
the IAC. We appreciate the work of Michael Leveille at STScI for facilitating the HST visits.
We also thank members of the.MOB (Months of Betelgeuse) for helpful and spirited discussion.
This work was supported in part by STScI Grant HST-G0-15641.001  to the Smithsonian Astrophysical Observatory.
MM acknowledges support from the ERC consolidator grant 646758 AEROSOL.

\vspace{5mm}
\facilities{HST(STIS); AAVSO; Izana(STELLA); FLWO (TRES)}

\end{document}